

\input harvmac

\overfullrule=0pt


\def\C{{\scriptscriptstyle C}}

\def\J{{\scriptscriptstyle J}}

\def\P{{\scriptscriptstyle P}}

\def\T{{\scriptscriptstyle T}}


\def\CA{{\cal A}}


\def\a{\alpha}
\def\b{\beta}
\def\d{\delta}

\def\g{\gamma}

\def\r{\rho}
\def\s{\sigma}
\def\t{\tau}
\def\th{\theta}
\def\u{\mu}
\def\v{\nu}
\def\z{\zeta}


\def\aS{\alpha_s}

\def\bbar{{\overline b}}
\def\Br{{\rm Br}}
\def\cbar{{\overline c}}
\def\ccdot{\hbox{\kern-.1em$\cdot$\kern-.1em}}
\def\chicJ{{\chi_{c\scriptscriptstyle J}}}
\def\chicJh{{\chi_{c\scriptscriptstyle J}^{(h)}}}
\def\chicJhabs{{\chi_{c\scriptscriptstyle J}^{(|h|)}}}
\def\chione{\chi_{c1}}
\def\chitwo{\chi_{c2}}
\def\chizero{\chi_{c0}}
\def\Dbar{{\overline D}}

\def\eh{\varepsilon^{(h)}}

\def\GeV{\>\, \rm GeV}

\def\gtap{\raise.3ex\hbox{$>$\kern-.75em\lower1ex\hbox{$\sim$}}}
\def\J{J/\Psi}
\def\ltap{\raise.3ex\hbox{$<$\kern-.75em\lower1ex\hbox{$\sim$}}}

\def\mc{m_c}

\def\MeV{\> {\rm MeV}}

\def\np{{n \ccdot p}}

\def\pbar{{\overline{p}}}

\def\pperp{p_\perp}
\def\Ptrans{P^\T}

\def\TeV{\>\, {\rm TeV}}


\def\half{{1 \over 2}}

\def\third{{1 \over 3}}
\def\threehalves{{3 \over 2}}
\def\threefourths{{3 \over 4}}

\def\twothirds{{2 \over 3}}


\newdimen\pmboffset
\pmboffset 0.022em
\def\oldpmb#1{\setbox0=\hbox{#1}%
 \copy0\kern-\wd0 \kern\pmboffset\raise
 1.732\pmboffset\copy0\kern-\wd0 \kern\pmboffset\box0}


%
%
\def\appendix#1#2{\global\meqno=1\global\subsecno=0\xdef\secsym{\hbox{#1.}}
\bigbreak\bigskip\noindent{\bf Appendix. #2}\message{(#1. #2)}
\writetoca{Appendix {#1.} {#2}}\par\nobreak\medskip\nobreak}

\nref\CDF{CDF Collaboration, Fermilab-conf-94/136-E (1994), unpublished.}
\nref\Braaten{E. Braaten, M. A. Doncheski, S. Fleming and M. L. Mangano,
 Fermilab-pub-94/135-T (1994), unpublished.}
\nref\Cacciari{M. Cacciari and M. Greco, FNT/T-94/13 (1994), unpublished.}
\nref\BraatenYuanI{E. Braaten and T.C. Yuan, Phys. Rev. Lett. {\bf 71} (1993)
 1673.}
\nref\BCY{E. Braaten, K. Cheung and T.C. Yuan, Phys. Rev. {\bf D48} (1993)
 4230.}
\nref\Chen{Y.-Q. Chen, Phys. Rev. {\bf D48} (1993) 5181.}
\nref\BraatenYuanII{E. Braaten and T.C. Yuan, Fermilab-pub-94/040-T (1994),
 unpublished.}
\nref\Falk{A. Falk, M. Luke, M. Savage and M. Wise, Phys. Lett {\bf B312}
 (1993) 486.}
\nref\Kuhn{J. H. K\"uhn, J. Kaplan and E. G. O. Safiani, Nucl. Phys. {\bf
 B157} (1979) 125\semi
 B. Guberina, J.H. K\"uhn, R.D. Peccei and R. R\"uckl, Nucl. Phys. {\bf B174}
 (1980) 317.}
\nref\CheungYuan{K. Cheung and T. C. Yuan, NUHEP-TH-94-7 (1994), unpublished.}
\nref\Mertig{R. Mertig, M. B\"ohm and A. Denner, Comp. Phys. Comm. {\bf 64}
 (1991) 345.}
\nref\Ma{J. P. Ma, Phys. Lett. {\bf B332} (1994) 398.}
\nref\Isgur{S. Godfrey and N. Isgur, Phys. Rev. {\bf D32} (1985) 189.}


\nfig\Pwavegraphs{Lowest order Feynman diagrams which mediate gluon
fragmentation to color-singlet P-wave charmonium bound states.}
\nfig\Swavegraph{Lowest order Feynman diagram which mediates gluon
fragmentation to a color-octet S-wave charmonium state.}


\def\CITTitle#1#2#3{\nopagenumbers\abstractfont
\hsize=\hstitle\rightline{#1}
\vskip 0.4in\centerline{\titlefont #2} \centerline{\titlefont #3}
\abstractfont\vskip .4in\pageno=0}


\CITTitle{{\baselineskip=12pt plus 1pt minus 1pt
  \vbox{\hbox{CALT-68-1943}\hbox{DOE RESEARCH AND}\hbox{DEVELOPMENT
  REPORT}\hbox{FERMILAB-PUB-94/256-T}}}}
{Gluon Fragmentation into Polarized Charmonium}{}
\centerline{
  Peter Cho\footnote{$^1$}{Work supported in part by an
  SSC Fellowship and by the U.S. Dept. of Energy under DOE Grant no.
  DE-FG03-92-ER40701.}
  and Mark B. Wise\footnote{$^2$}{Work supported in part by
  the U.S. Dept. of Energy under DOE Grant no. DE-FG03-92-ER40701.}}

\centerline{Lauritsen Laboratory}
\centerline{California Institute of Technology}
\centerline{Pasadena, CA  91125}
\medskip\centerline{and}\medskip
\centerline{Sandip P. Trivedi\footnote{$^3$}{Work supported in part by DOE
Contract no. DE-AC02-76CHO3000.}}
\centerline{Fermi National Accelerator Laboratory}
\centerline{P.O. Box 500, Batavia, IL 60510}

\vskip .2in
\centerline{\bf Abstract}
\bigskip

	Gluon fragmentation to $\chicJ(1P)$ followed by single photon
emission represents the dominant source of prompt $\J$'s at the
Tevatron for $\pperp \gtap 6 \GeV$.  Since fragmenting gluons are
approximately transverse, their products are significantly polarized.
We find that gluon fragmentation populates the helicity levels of
$\chione$, $\chitwo$ and $\J$ according to $D_{\chione^{(h=0)}}:
D_{\chione^{(|h|=1)}} \simeq 1:1$, $D_{\chitwo^{(h=0)}} :
D_{\chitwo^{(|h|=1)}} : D_{\chitwo^{(|h|=2)}} \simeq 1:2.9:6.0$ and
$D_{\J^{(h=0)}} : D_{\J^{(|h|=1)}} \simeq 1:3.4$.  We also speculate
that gluon fragmentation to the radially excited $\chitwo(2P)$ state
followed by subsequent radiative decay could represent a large source
of $\psi'(2S)$'s and potentially resolve the $\psi'$ deficit problem.
A measurement of these states' polarizations would test this idea.

\Date{8/94}


	The production of the $\J$ charmonium bound state is currently
under active study at Fermilab \CDF.  Until recently, the dominant
sources of $\J$'s at a hadron collider were believed to be parton
fusion and $B$ meson decay.  These two processes respectively produce
prompt and delayed $\J$'s which can be distinguished via $B$ meson
vertex displacement measurements.  Comparison between the theoretical
prediction and experimental measurement of the transverse momentum
differential cross section $d\sigma(p\pbar\to \J+X)/d\pperp$ section
reveals that parton fusion alone underestimates the prompt $\J$
production rate at high transverse momenta by approximately an order
of magnitude \refs{\Braaten,\Cacciari}.  Such a large discrepancy
between theory and data clearly indicates that another prompt
production mechanism must be at work.

	Within the past few years, parton fragmentation has been
examined as an alternate source of $\J$'s \BraatenYuanI.  Although
fragmentation takes place at higher order in perturbative QCD than
quark or gluon fusion, the falloff of the former with increasing
transverse momentum is much slower than that of the latter.  So for
$\pperp \gtap 6 \GeV$, parton fragmentation represents the dominant
source of prompt $\J$'s.

	The first charmonium fragmentation functions to be calculated
were $D_{g \to \J}(z)$ and $D_{c \to \J}(z)$ which specify the
probability for gluons and charm quarks to hadronize into $\J$ as a
function of its longitudinal momentum fraction $z$
\refs{\BraatenYuanI{--}\Chen}.  The only nonperturbative piece of
information which enters into the lowest order computation of these
$S$-wave fragmentation functions is the square of the charmonium bound
state's wavefunction at the origin.  The remainder of the calculation
is based upon perturbative QCD.  More recently, the fragmentation
functions for gluons and charm quarks to hadronize into the lowest
lying $P$-wave charmonium bound states $\chizero$, $\chione$ and
$\chitwo$ have been computed \BraatenYuanII.  These $\chicJ$ states
radiatively decay down to $\J$ with the branching ratios 0.7\%, 27\%
and 14\% for $J=0$, 1 and 2 respectively.  After folding together
these branching ratios with the $P$-wave fragmentation functions, one
finds that gluon fragmentation to $\chicJ$ followed by single photon
emission to $\J$ dominates at high $\pperp$ over all other prompt
mechanisms by more than an order of magnitude.  When this $\J$ source
is included, the theoretical prediction for $d\sigma(p\pbar \to \J +
X)/d\pperp$ at $\sqrt{s} = 1.8 \TeV$ moves to within a factor of two
of recent CDF data.

	Most of the fragmentation functions which have been calculated
to date describe the production of unpolarized quarkonium.  However,
it is straightforward to compute polarized fragmentation functions as
well.  Charm fragmentation into transverse and longitudinal $\J$'s was
considered in refs.~\Chen\ and \Falk\ and found to yield essentially
no polarization.  $\J$'s produced at a lepton collider like LEP are
therefore not expected to be significantly polarized.  But since gluon
fragmentation to $\chicJ$ represents the dominant source of $\J$'s at
a hadron machine like the Tevatron, the polarized fragmentation
functions $D_{g \to \chicJhabs}(z)$ where $|h| \le J$ denotes the
helicity of the produced $\chicJ$ must be determined before the degree
of $\J$ polarization can be estimated.  We present the results for
these fragmentation functions and the $\chicJ$ and $\J$ polarizations
which they induce in this letter.

	To begin, we adopt the notation and general methods for
computing $P$-wave fragmentation functions established in
\BraatenYuanII.  The lowest order diagrams that contribute to $\chicJ$
fragmentation are illustrated in \Pwavegraphs.  They may be evaluated
using standard Feynman rules for quarkonium processes \Kuhn.  The
kinematic regime in which these graphs become important occurs when
the lab frame energy $q_0$ of the incoming off-shell gluon $g^*$ is
large, but its squared four-momentum $q^2$ is close to the square of
the charmonium bound state's mass.  We therefore neglect terms which
are subdominant in the ratio $q^2 / q_0^2$.  The sum of the two
diagrams in \Pwavegraphs\ logarithmically diverges in the limit $z \to
1$ when the bound state carries off all the original gluon's energy
and the outgoing gluon undergoes zero recoil.  This infrared
divergence is canceled by the diagram in \Swavegraph\ which depicts
the conversion of $g^*$ into a color-octet $^3S_1$ state.  The colored
state can subsequently emit a soft gluon and decay to a color-singlet
$\chicJ$.  The graphs in figs.~1 and 2 must be added together to
obtain an infrared finite result.

	To determine the polarized fragmentation functions $D_{g \to
\chicJhabs} (z)$, we need the polarization sums for the individual
helicity levels of $\chicJ$.  These spin sums may be conveniently
expressed in terms of an auxiliary light-like vector $n=(1,-\hat{p})$
where $\hat{p}$ denotes a unit vector oriented along the
three-momentum of the $\chicJ$ in the lab frame.  The longitudinal and
transverse polarization sums for the spin-1 boson can then be simply
written in the covariant forms \CheungYuan\
\eqn\spinsumsI{\eqalign{
\sum_{h=0} \eh_\a(p) \, \eh_\b(p)^* &= P_{\a\b} - \Ptrans_{\a\b} \cr
\sum_{|h|=1} \eh_\a(p) \, \eh_\b(p)^* &= \Ptrans_{\a\b} \cr}}
where
\eqn\projectors{\eqalign{
\Ptrans_{\a\b} &= -g_{\a\b} + {1 \over \np} \bigl(p_\a n_\b + n_\a p_\b
 \bigr) - {p^2 \over (\np)^2} n_\a n_\b \cr
P_{\a\b} &= -g_{\a\b} + {p_\a p_\b \over p^2} \cr}}
represent two-dimensional transverse and three-dimensional projection
operators.  The corresponding spin-2 polarization sums are given by
\eqn\spinsumsII{\eqalign{
\sum_{h=0} \eh_{\a\b}(p) \, \eh_{\u\v}(p)^* &= \twothirds \bigl(P_{\a\b} -
 \threehalves \Ptrans_{\a\b} \bigr) \bigl( P_{\u\v} - \threehalves
\Ptrans_{\u\v} \bigr) \cr
\sum_{|h|=1} \eh_{\a\b}(p) \, \eh_{\u\v}(p)^* &= \half \Bigl[ P_{\a\u} P_{\b\v}
 + P_{\a\v} P_{\b\u} - \Ptrans_{\a\u} \Ptrans_{\b\v} - \Ptrans_{\a\v}
 \Ptrans_{\b\u} \Bigr] \cr
 & \quad + \Bigl[ P_{\a\b} \Ptrans_{\u\v} + \Ptrans_{\a\b} P_{\u\v} -
 P_{\a\b} P_{\u\v} - \Ptrans_{\a\b} \Ptrans_{\u\v} \Bigr] \cr
\sum_{|h|=2} \eh_{\a\b}(p) \, \eh_{\u\v}(p)^* &= \half \Bigl[ \Ptrans_{\a\u}
 \Ptrans_{\b\v} + \Ptrans_{\a\v} \Ptrans_{\b\u} - \Ptrans_{\a\b}
 \Ptrans_{\u\v} \Bigr]. \cr}}

	After a straightforward computation,
\foot{We used the high energy physics Mathematica package Feyncalc to perform
much of the tedious algebra \Mertig.}
we obtain the polarized $\chicJ$ fragmentation  functions:
\eqn\fragzero{\eqalign{
D_{g \to \chizero}(z,M) & = {4 \over 81} {H_1 \aS(M)^2 \over M} \Bigl[
 {1 \over (1-z)_+} + \Bigl({13 \over 12} - \log{2\Lambda\over M}\Bigr) \d(1-z)
 -1 + {85 \over 8} z - {13 \over 4}z^2 \cr
 & \quad + {9 \over 4} (5-3z) \log(1-z) \Bigr]
 + {1 \over 12} {\pi \aS(M) H'_8(\Lambda) \over M} \d(1-z) \cr}}
\eqna\fragone
$$ \eqalignno{D_{g \to \chione^{(h=0)}}(z,M) &=
 {4 \over 81} {H_1 \aS(M)^2 \over M} \Bigl[
 {3 \over 2 (1-z)_+} + \Bigl({13 \over 8} -
 \threehalves\log{2\Lambda\over M}\Bigr) \d(1-z) - \threehalves
 - \threefourths z - \threehalves z^2 \Bigr] \cr
 & \quad + {1 \over 8} {\pi \aS(M) H'_8(\Lambda) \over M} \d(1-z)
 & \fragone a \cr
D_{g \to \chione^{(|h|=1)}}(z,M) &=
 {4 \over 81} {H_1 \aS(M)^2 \over M} \Bigl[
 {3 \over 2 (1-z)_+} + \Bigl({5 \over 4} -
 \threehalves \log{2\Lambda\over M}\Bigr) \d(1-z) - \threehalves -
 \threehalves z^2 \Bigr] \cr
 & \quad + {1 \over 8} {\pi \aS(M) H'_8(\Lambda) \over M} \d(1-z)
 & \fragone b \cr} $$
\eqna\fragtwo
$$ \eqalignno{D_{g \to \chitwo^{(h=0)}}(z,M) &=
 {4 \over 81} {H_1 \aS(M)^2 \over M} \Bigl[
 {1 \over 2 (1-z)_+} + \Bigl({13 \over 24} -
 \half\log{2\Lambda\over M}\Bigr) \d(1-z) \cr
 & \quad + 108 z^{-3} - 216 z^{-2}
 + 117 z^{-1} - {19 \over 2} - {5 \over 4} z - \half  z^2 \cr
 & \quad + 54 {(2-z)(1-z)^2 \over z^4} \log(1-z) \Bigr]
 + {1 \over 24} {\pi \aS(M) H'_8(\Lambda) \over M} \d(1-z)
 & \fragtwo a \cr
 & \cr
D_{g \to \chitwo^{(|h|=1)}}(z,M) &=
 {4 \over 81} {H_1 \aS(M)^2 \over M} \Bigl[
 {3 \over 2 (1-z)_+} + \Bigl({5 \over 4} -
 \threehalves\log{2\Lambda\over M}\Bigr) \d(1-z) \cr
 & \quad - 144 z^{-3} + 288 z^{-2}
 - 228 z^{-1} + {165 \over 2} - 6 z - \threehalves z^2 \cr
 & \quad - 36 {(2-z)(1-z)\over z^4} (z^2 - 2 z + 2) \log(1-z) \Bigr]
 + {1 \over 8} {\pi \aS(M) H'_8(\Lambda) \over M} \d(1-z) \cr
 && \fragtwo b \cr
D_{g \to \chitwo^{(|h|=2)}}(z,M) &=
 {4 \over 81} {H_1 \aS(M)^2 \over M} \Bigl[
 {3 \over (1-z)_+} + \Bigl({13 \over 4} -
 3 \log{2\Lambda\over M}\Bigr) \d(1-z) \cr
 & \quad + 36 z^{-3} - 72 z^{-2}
 + 111 z^{-1} - 78 + 21 z - 3 z^2 \cr
 & \quad + 9 {(2-z)\over z^4} (z^4 - 4 z^3 + 6 z^2 - 4 z + 2) \log(1-z) \Bigr]
 + {1 \over 4} {\pi \aS(M) H'_8(\Lambda) \over M} \d(1-z). \cr
 && \fragtwo c \cr} $$
In these expressions, $M \simeq 2 \mc$ denotes the charmonium bound
state's mass, $\Lambda$ represents the infrared cutoff, and $H_1 = 72
|R'_1(0)|^2/(\pi M^4)$ and $H'_8(\Lambda) = 8 |R^{(8)}_0(0)|^2/(3\pi
M^2)$ respectively contain the squares of the derivative of the
color-singlet $P$-wave and color-octet $S$-wave radial wavefunctions.
If the polarized fragmentation functions are summed over their
helicity levels, we recover the unpolarized $P$-wave fragmentation
functions reported in ref.~\BraatenYuanII.
\foot{The longitudinal and transverse $\chione$ fragmentation functions were
calculated in ref.~\Ma.  Our results in eqns.~\fragone{a}\ and
\fragone{b}\ agree with those in \Ma\ for $z \ne 1$ but differ for
$z=1$.}

	We adopt the parameter values $M=3500 \MeV$ and $\aS(M)=0.24$.
Following \BraatenYuanII, we also set the infrared cutoff to
$\Lambda=M/2$ and take $H_1 \simeq 15 \MeV$ and $H'_8(M/2) \simeq
3\MeV$.  The functions in eqns.~\fragzero\ - \fragtwo{}\ can be
evolved from the charmonium scale to higher energies using the
Altarelli-Parisi equation and folded together with the gluon cross
section $d\sigma(p \pbar \to g+X)/d\pperp$ to obtain the transverse
momentum distribution of $\chicJ$'s produced at the Tevatron:
\eqn\chicJxsect{{d\sigma(p\pbar \to \chicJhabs+X)\over d\pperp} = \int_0^1 dz
{d\sigma(p\pbar \to g({\pperp / z}) +X,\mu) \over d\pperp}
D_{g\to \chicJhabs}(z,\mu).}
Since the gluon cross section is a very rapidly decreasing function of
$\pperp$, it is a good approximation to evaluate the integral in
\chicJxsect\ retaining just the terms proportional to $\d(1-z)$ in the
fragmentation functions:
\foot{Altarelli-Parisi evolution approximately cancels in the ratios of
fragmentation functions.  We therefore neglect it in our polarization
analysis.}
\eqn\fragprob{\eqalign{
D_{g \to \chizero}(z,M) &\simeq 0.66 \times 10^{-4} \d(1-z) + \cdots
\cr D_{g \to \chione^{(h=0)}}(z,M) &\simeq 1.00 \times 10^{-4} \d(1-z)
+ \cdots \cr D_{g \to \chione^{(|h|=1)}}(z,M) &\simeq 0.95 \times
10^{-4} \d(1-z) + \cdots \cr D_{g \to \chitwo^{(h=0)}}(z,M) &\simeq
0.33 \times 10^{-4} \d(1-z) +\cdots \cr D_{g \to
\chitwo^{(|h|=1)}}(z,M) &\simeq 0.95 \times 10^{-4} \d(1-z) + \cdots
\cr D_{g \to \chitwo^{(|h|=2)}}(z,M) &\simeq 1.99 \times 10^{-4}
\d(1-z) + \cdots.\cr}}
Longitudinally and transversely polarized $\chione$'s are therefore
produced at the Tevatron in the ratio $D_{\chione^{(h=0)}} : D_{
\chione^{(|h|=1)}} \simeq 1:1$, while the helicity levels of $\chitwo$ are
populated according to $D_{\chitwo^{(h=0)}} : D_{\chitwo^{(|h|=1)}} :
D_{\chitwo^{(|h|=2)}} \simeq 1:2.9:6.0$.  Comparing these $J=1$ and
$J=2$ ratios to their unpolarized counterparts $D_{\chione^{(h=0)}} :
D_{\chione^{(|h|=1)}} = 1:2$ and $D_{\chitwo^{(h=0)}} :
D_{\chitwo^{(|h|=1)}} : D_{\chitwo^{(|h|=2)}} = 1:2:2$, we clearly see
that the $\chione$ and $\chitwo$ states produced as a result of gluon
fragmentation are significantly polarized.

	The source of this sizable $\chicJ$ polarization can be traced
to the fragmenting gluon.  If the gluon were on-shell, its
polarization would be completely transverse.  The extent to which
$g^*$ is actually off-shell modifies this result by only
$O(q^2/q_0^2)$ terms.  This effect can be seen most simply in the
diagram of \Swavegraph.  The color-octet $^3S_1$ $c\cbar$ in the
$|c\cbar g\rangle$ Fock state must inherit the gluon's polarization in
order to conserve angular momentum.  The subsequent transformation of
this state into $\chione$ and $\chitwo$ populates their respective
$|h|=0,1$ and $|h|=0,1,2$ helicity levels in the ratios 1:1 and 1:3:6.
As the color-octet terms in eqns.~\fragzero\ - \fragtwo{}\ numerically
dominate over the color-singlet terms, this explanation accounts in
large part for the $\chicJ$ polarization which we have found.

	We now turn to consider the radiative decay $\chicJ \to \J +
\gamma$.  Since the electromagnetic branching ratio for $\chizero$ is
more than an order of magnitude smaller than those for $\chione$ and
$\chitwo$, we neglect its contribution to $\J$ production.  The
invariant amplitudes for the remaining $J^{\P\C}=(1,2)^{++}$ $P$-wave
charmonium states to decay via E1 transitions to the
$J^{\P\C}=(1)^{--}$ $S$-wave state must be parity even, charge
conjugation symmetric and electromagnetic gauge invariant.  They can
be written down by inspection:
\eqn\amps{\eqalign{
i\CA(\chione(p) \to \J(p-k) + \gamma(k)) &= g_1 \epsilon^{\u\v\a\b} k_\u
 \varepsilon^{(\chione)}_\v \varepsilon^{(\J)}_\a
 \varepsilon^{(\gamma)}_\b \cr
i\CA(\chitwo(p) \to \J(p-k) + \gamma(k)) &= g_2 p^\u
 \varepsilon_{(\chitwo)}^{\a\b}
 \varepsilon^{(\J)}_\a
 \bigl[ k_\u \varepsilon^{(\gamma)}_\b - k_\b \varepsilon^{(\gamma)}_\u \bigr].
 \cr}}

	Given these amplitudes, we can determine the photon's angular
distribution in the $\chicJ$ rest frame.  Letting $\th$ denote the
angle between the photon's three-momentum in this frame and the $\chicJ$'s
three-momentum in the lab frame, we form the dimensionless ratio
\eqn\ratio{
R^{(J)}(\cos\th) =
%
\sum_{h=-J}^J {D_{g \to \chicJh} \over D_{g \to \chicJ}}
 {\displaystyle{d\Gamma\over d\cos\th}
 (\chicJh \to \J + \gamma) \over
 \Gamma(\chicJ \to \J+\gamma)} }
which is a convenient measure of the angular distribution of photons
from polarized $\chicJ$'s.  The explicit dependence of $R^{(J)}$ upon
$\cos\th$ is given by
\eqn\Ronetwo{\eqalign{
R^{(1)}(\cos\th) &= {3\over 8} \bigl[ \bigr(1+{\r\over 2}\bigl)
 + \bigr(1-\threehalves \r\bigr) \cos^2\th\bigr] \cr
R^{(2)}(\cos\th) &= \threefourths\bigl[ \bigl({5 \over 6} - {\s\over 12}
 - {\t\over 3} \bigr) - \bigl(\half-{\s\over 4}-\t\bigr) \cos^2\th \bigr]}}
where $\r=D_{g \to \chione^{(|h|=1)}}/D_{g \to \chione}$,
$\s=D_{g\to\chitwo^{(|h|=1)}}/ D_{g\to\chitwo}$ and
$\t=D_{g\to\chitwo^{(|h|=2)}}/D_{g\to\chitwo}$.  If $\chione$ and
$\chitwo$ were unpolarized, then $\r=2/3$, $\s=\t=2/5$ and $R^{(J)}$
would become independent of $\cos\th$.  But the fragmentation results
in eqn.~\fragprob{}\ imply $\r\simeq 0.49$, $\s\simeq 0.29$ and
$\t\simeq 0.61$ and yield
\eqna\Rvalues
$$ \eqalignno{R^{(1)}(\cos\th) &\simeq 0.47 \bigl[1+0.21 \cos^2 \th\bigr] &
\Rvalues a\cr
R^{(2)}(\cos\th) &\simeq 0.46 \bigl[1 + 0.30 \cos^2\th\bigr].
& \Rvalues b \cr}$$
In principle, measurements of $R^{(1)}$ and $R^{(2)}$ would determine
the polarizations of $\chione$ and $\chitwo$.  But in practice, it
will be much easier to observe the average angular distribution
\eqn\Ravg{R^{\rm (avg)}(\cos\th) =
{\displaystyle{\sum_{J=1}^2} D_{g \to \chicJ}
 \Br(\chicJ \to \J+\g) R^{(J)}(\cos\th) \over
\displaystyle{\sum_{J=1}^2} D_{g \to \chicJ} \Br(\chicJ \to\J+\g)}
\simeq 0.47 \bigr[1+0.25 \cos^2\th\bigl]}
and extract an average $\chicJ$ polarization.

	The amplitude expressions in \amps\ can also be used to derive
the polarization of the $\J$ which is induced by its $\chicJ$
progenitor.  The feeddown from the separate $\chicJ$ helicity modes to
those of the $\J$ is given by
\eqn\feeddown{\eqalign{
D_{\J^{(h=0)}} &= \Br(\chione\to\J+\gamma)
\Bigl[\half D_{\chione^{(|h|=1)}}\Bigr]
+\Br(\chitwo\to\J+\gamma) \Bigl[\twothirds D_{\chitwo^{(h=0)}} + \half
 D_{\chitwo^{(|h|=1)}} \Bigr] \cr
D_{\J^{(|h|=1)}} &= \Br(\chione\to\J+\gamma) \Bigl[ D_{\chione^{(h=0)}} +
\half D_{\chione^{(|h|=1)}} \Bigr] \cr
&+ \Br(\chitwo\to\J+\gamma) \Bigl[\third D_{\chitwo^{(h=0)}} + \half
 D_{\chitwo^{(|h|=1)}} + D_{\chitwo^{(|h|=2)}}\Bigr]. \cr}}
After inserting the radiative branching ratios and $\chicJ$
fragmentation probabilities, we find that longitudinal and transverse
$\J$'s are produced in the ratio $D_{\J^{(h=0)}} : D_{\J^{(|h|=1)}}
\simeq 1:3.4$.  Equivalently, the ratio of transversely polarized to
total $\J$'s equals $\z \simeq 0.77$.  This ratio may be measured in
the leptonic decay $\J \to
\ell^+ \ell^-$.  If $\Theta$ represents the angle between the lepton
three-momentum in the $\J$ rest frame and the three-momentum of the
$\J$ in the lab frame, then
\eqn\lepton{{\displaystyle{d\Gamma \over d\cos\Theta} (\psi\to\ell^+\ell^-)
\over\Gamma(\psi \to \ell^+ \ell^-)} = \threefourths \Bigl[\Bigl(1-{\z\over 2}
\Bigr) - \Bigl(1-\threehalves\z\Bigr) \cos^2\Theta \Bigr] \simeq
0.46 \bigl[1 + 0.25 \cos^2\Theta\bigr].}
$\J$ production via gluon fragmentation to $\chicJ$ consequently
induces a 25\% shift in the lepton pair angular distribution relative
to unpolarized $\J$'s.

	The $\chicJ$ and $\J$ polarizations that we have found
represent model independent predictions of QCD which can be
experimentally tested.  Verification of these results would provide
nontrivial checks of the entire fragmentation picture of quarkonium
production at large $\pperp$.

	We have so far considered the production of only the lowest
lying $n=1$ radial level charmonium states.  However, fragmentation
ideas can be simply applied to higher radial levels as well.  In
particular, gluon and charm fragmentation to $\psi'(2S)$ have been
studied in ref.~\Braaten.  Even when fragmentation is included along
with direct production, the theoretical prediction for $\psi'$
production underestimates the number of $\psi'$'s which have been
observed at the Tevatron by roughly a factor of 50.  This large gap
between theory and data strongly suggests that some important $\psi'$
production mechanism still remains to be included.

	It is important to recall that $\psi'$ is the heaviest
$c\cbar$ bound state which lies below the $D\Dbar$ threshold.
Therefore, $n=1$ $\chicJ$ states cannot radiatively decay to $\psi'$,
but their $n=2$ counterparts can.  None of these latter states which
lie above the $D\Dbar$ threshold have been observed.  Estimates for
their masses yield $M(\chizero(2P))=3920 \MeV$, $M(\chione(2P))=3950
\MeV$ and $M(\chitwo(2P))=3980 \MeV$ \Isgur.  These mass values kinematically
allow the $S$-wave transitions $\chizero(2P) \to D\Dbar$ and
$\chione(2P)\to D^* \Dbar$ to occur.  We therefore expect the $J=0$ and
$J=1$ excited $\chicJ$'s to be broad and to have negligible branching
fractions to lower $c\cbar$ bound states.  However, angular momentum
and parity considerations require the analogous decays $\chitwo(2P)
\to D\Dbar$ and $\chitwo(2P) \to D^* \Dbar$ for the $J=2$ state to
proceed via $L=2$ partial waves.  Although we cannot readily compute
by how much these $D$-wave decays will be suppressed, it is possible
that the branching fractions for $\chitwo(2P)$ transitions to
charmonium states below $D\Dbar$ threshold could be significant.
\foot{The $^1D_2$ charmonium state is forbidden from decaying to $D\Dbar$
by parity.  Moreover, its mass is predicted to lie below the $D^*
\Dbar$ threshold \Isgur.  This state is therefore narrow.  However, its
production is suppressed, and it is expected to have a very small
branching ratio to $\psi' \gamma$.  The contribution of the $^1D_2$
state to $\psi'$ production is consequently negligible.}
If so, the experimentally measured branching fraction
$\Br(\chi_{b2}(2P) \to \Upsilon(2S) + \gamma)=19\%$ in the $b\bbar$
sector suggests that the corresponding fraction $\Br(\chitwo(2P) \to
\psi'(2S)+\gamma)$ could lie in the few percent range.  We estimate
that a 5\% branching fraction would enhance the theoretical prediction
for $\psi'$ production by more than an order of magnitude.  Such an
enhancement would help resolve the $\psi'$ deficit problem.

	Our proposal is admittedly speculative.  If this idea is
correct, then a $\psi'$ should be accompanied by a photon resulting
from $\chitwo(2P)$ decay.  Since the $\chitwo(2P)$ is polarized to
approximately the same extent as its $n=1$ counterpart, the photon
will be distributed in angle according to $R^{(2)}(\cos\th)$ as
specified in eqn.~\Rvalues{b}.  Moreover, the induced polarization for
$\psi'$ should be slightly enhanced relative to that of $\J$ since
$\chione(2P)$ does not feed down along with $\chitwo(2P)$.  A
measurement of these radially excited states' polarizations would
therefore provide a test of this possible $\psi'$ production
mechanism.

\listrefs
\listfigs
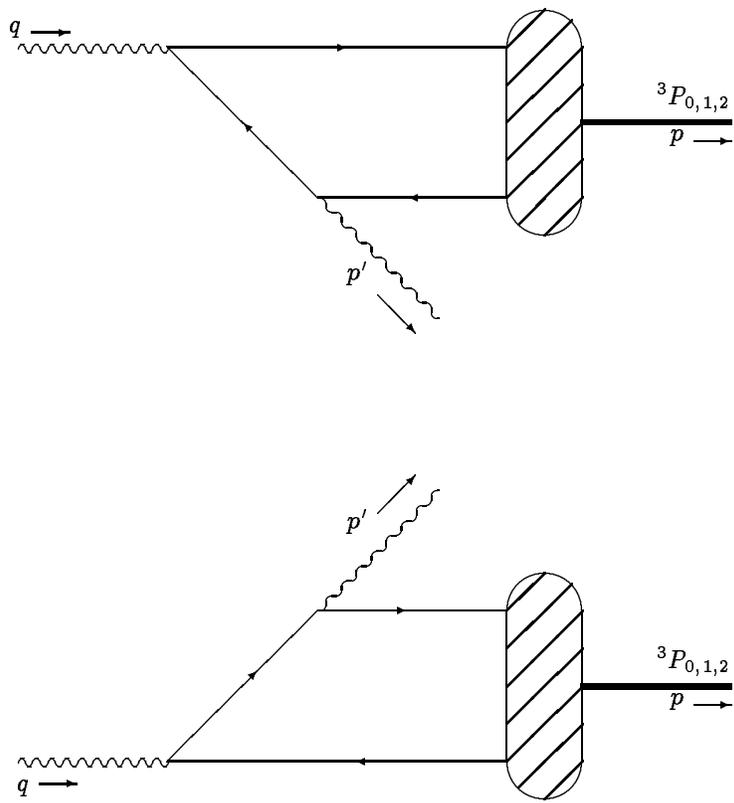
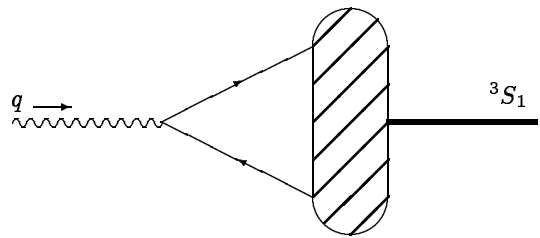
\bye